\documentclass{icrc2009}

\usepackage{graphicx}   % for including figures
\usepackage{caption}    % for captions
\usepackage[font=footnotesize]{subfig} % subfig.sty for a double column floating figure using two subfigures
\usepackage{fixltx2e}
\usepackage{url}

\newcommand{\shorttitle}[1]%
{\markboth{Proceedings of the 31\MakeLowercase{$^{st}$} ICRC, {\L}\'{o}d\'{z} 2009}{#1} }
\begin{document}
\title{$\gamma$-ray astronomy in the summer of 2009}

\author{\IEEEauthorblockN{Diego F. Torres\IEEEauthorrefmark{1}
			}
                            \\
\IEEEauthorblockA{\IEEEauthorrefmark{1}ICREA \& Institut de Ci\`encies de l'Espai (IEEC-CSIC) \\
Campus UAB, Fac. de Ci\`encies, Torre C5, parell, 2a planta
08193 Barcelona,  Spain}
}

% please write the preseter's name and short title (3-4 words maximum)
%    which will appear at the header of the even pages.
\shorttitle{D. F. Torres, $\gamma$-ray astronomy in the summer of 2009}
\maketitle

\begin{abstract}
 This paper presents a snapshot of the field of $\gamma$-ray  astrophysics in the early summer of 2009, as it was discussed in about 200 presentations at the International Cosmic Ray Conference (ICRC) held in {\L}\'{o}d\'{z}, Poland. This is the written wrap up of a Rapporteur, one-hour talk and as such it is thus an atypical review, a still picture in a moment of great advancement in an observationally driven field, led by the full operation of ground-based arrays and the launch and operations of Fermi and Agile. 
  \end{abstract}

\begin{IEEEkeywords}
$\gamma$-ray  astronomy
\end{IEEEkeywords}
 
\section{Introduction}

This paper intends to provide a summary of the presentations, posters, and talks in the field of $\gamma$-ray  astronomy that have been submitted to the International Cosmic Ray Conference (ICRC) held in {\L}\'{o}d\'{z}, Poland, in the early summer of 2009. Thus, it is not a strict review, but rather a still picture of the field at the moment of the meeting, with only minimal excursions into literature not therein discussed. This is the written wrap up of a Rapporteur, one-hour talk, which although perhaps more complete than that, it is probably not fully comprehensive and completely devoid of oversights: In total, this summary intends to select and present some of the most important results discussed in 13 papers submitted to the Diffuse, 83 papers to the Galactic, 60 to the Extragalactic, and 28 to the GRB sessions. 

One of the most remarkable events of the 2009 ICRC is that this meeting has been the first in which Fermi and Agile, the two currently flying GeV satellites, are fully operational. The number and interest of their results has been impressive, as one can see along this summary.  In addition, H.E.S.S., VERITAS, and MAGIC collaborations continue to lead the ground-based arrays in producing first-rate science, with amazing discoveries reported at the meeting, some long-time expected, some not. Finally, MILAGRO reported on its survey of the   
the higher energy sky, finding puzzling coincidences across all energy regimes.

\section{Diffuse emission} 
  
The diffuse Galactic  $\gamma$-ray emission is produced by cosmic rays (CRs) protons, nuclei, and electrons 
interacting
with the interstellar gas and the radiation fields.
Measurements by the Energetic Gamma-Ray Experiment Telescope (EGRET) 
instrument on the  Compton Gamma-Ray Observatory (CGRO) early indicated an
excess of  $\gamma$-ray emission at energies $\geq 1$ GeV relative to 
diffuse Galactic  $\gamma$-ray emission models that were 
consistent with directly measured CR spectra in the Earth neighborhood 
(this was dubbed the ``EGRET GeV excess''). The excess emission was observed in all directions on the sky, and because of that,  
universal explanations have been proposed, 
including beyond standard particle interaction models scenarios like annihilating or decaying 
dark matter. 

Porter et al. and Strong et al., both on behalf of the Fermi collaboration, presented the results that the Large Area Telescope onboard Fermi (see Atwood et al. 2009 for details) obtained 
about the measurements of the Galactic diffuse emission (for energies 100 MeV to 10 GeV and Galactic latitudes
$10^\circ \leq |b| \leq 20^\circ$).  Essentially, and perhaps this is one of the most notable results of the ICRC: the GeV bump 
earlier discovered by the EGRET experiment does 
not seem to exist when observed with Fermi. 
The intensity scales of the LAT and EGRET measurements have been found to be different, with
the result that the LAT-measured spectra are softer. The spectral shape is now roughly compatible with that of a model (not tuned to LAT data)  that 
is consistent with directly measured CR spectra,  i.e., no extra component is needed to explain any of the observations. 

One of the uncertainties of any model of the diffuse emission is the conversion between CO and H2, what is known as the X-factor. Is it the same across  the galaxy?   The ICRC featured three presentations on this topic, by Okumura et al. (Fermi), 
Tibaldo et al. (Fermi), and 
Giuliani et al. (Agile). The cases of the giant molecular clouds (GMC) complexes in Orion, presented by Okumura et al., serve as an example of the different analysis presented. Orion A and B are two of the brightest 
giant molecular clouds in diffuse MeV--GeV gamma 
rays, located at about 500 pc from Earth, with masses of the order of 10$^5 M_\odot$. 
No bright point 
sources of gamma rays are known in the Orion A and 
B regions, what makes them a good target for diffuse emission studies. In order 
to extract the diffuse emission, a mixture between modeling and measurement is made to separate the following components:
(A) pion gamma and CR electron bremsstrahlung from 
GMC (H2 region). 
(B) pion gamma and CR electron bremsstrahlung from 
H I and H II gas. 
(C) Inverse Compton (IC) scattering of CR electrons 
off interstellar radiation fields. 
(D) Extragalactic diffuse emission and unresolved weak 
point sources. 
(E) Instrumental residual background from misclassified interactions of cosmic rays in the LAT.
 The bremsstrahlung contribution to B is subject to relatively large (several tens 
of percent) uncertainty. However its contribution is important only in the energy range below 
100 MeV. Modeling allows then to subtract components B and C, 
and estimations of D and E are made from the residual in the off-source regions that are more than 3 degrees away from the 
clouds themselves and other bright point sources, assuming 
that these two components distribute uniformly. Once these components are subtracted, what is left is just the pion gamma and CR electron bremsstrahlung from 
GMCs (H2 region) and a map of it can be compared directly with the predictions (the $\gamma$-ray  emission) coming from the application of a given X-factor to a measured map of CO emission. 
By applying this method, 
it was found that 
the $\gamma$-ray  intensity 
map and that of CO map in the regions of the clouds 
correlates well, although the slopes relating the two fluxes of two regions 
are found to be different by a factor of 1.3. This may indicate that either the X factor is different in each of the cloud, or the cosmic ray density is. Differences are visible even within the same complex. 
Similar studies in other regions have been performed, and variance in the X factor is suggested as common. 

Ackerman et al. (Fermi) presented the LAT results for the observations of the extragalactic diffuse continuum $\gamma$-ray  emission.
 The diffuse emission observed by the Fermi LAT 
is also a superposition of several components,
with the highly 
structured Galactic component  being clearly dominant at low Galactic latitudes. The subtraction of this foreground is the biggest challenge for the determination of the extragalactic background. 
But based on careful modeling of the detected LAT point sources (which need to be subtracted too) and the 
 diffuse foreground, the extragalactic contribution can be isolated. 
 The results indicate an isotropic 
 diffuse emission that can be fitted by a power-law with index $-2.45$ between 200 MeV and 50 GeV.
 
 It is interesting to note that the isotropic diffuse flux contribution from unresolved sources depends on the LAT point source sensitivity. Because of that, the 
LAT isotropic flux is naturally expected to be below the EGRET level (factor $>$10 improvement in point source sensitivity), and furthermore, the 
contribution is expected to decrease with the accumulation of LAT observation time.
 
ICRC featured two presentation on the Large Magellanic Cloud (LMC), by the H.E.S.S. and the Fermi experiments.
At VHE, Komin et al. (H.E.S.S.) reported that the LMC is 
detected as a point-like source, 
consistent with SNR 157B/PSR J0537-6910 (which is the highest $\dot E$ pulsar, with a required efficiency of 0.01 in order to be compatible with the TeV observations). This would make of it the most distant VHE PWN detected. Komin et al. also reported an upper limit on SN1987A
and no detection from 30 Doradus (but still consistent with pion decay processes therein). This
VHE source is correlated with a Fermi bright source list detection, and within the LMC Fermi detection as reported by Porter et al.
In fact, the LAT was able to resolve a $\gamma$-ray bright region within the LMC LMC for the first time, with the maximum of the emission observed (but not all of the emission) 
coming from 30 Doradus. Without doubt, subsequent detailed papers from both collaborations will follow these preliminary presentations.

Finally, two presentations from VERITAS and H.E.S.S., by Benbow et al. and Domainko et al., respectively, informed
about the detection of the two nearest starburst galaxies at VHE. 
VERITAS observed the M82 galaxy along 
2007-2009 for about 137 h live time. Only dark time (no moonlight) was used in order to detect it, and finally appears as a point-like
5.0$\sigma$ excess (pre-trials), 4.8$\sigma$ (post-trials) at 
$E > 700$ GeV. This makes of M82 one of the
weakest VHE sources observed, with a luminosity about $0.9\%$ Crab. From 875 GeV to 5 TeV, a power-law fit well the spectrum (although the index is not well constrained yet), -2.5 $\pm$ 0.6. Both flux and spectrum are  consistent with previous predictions (e.g, de Cea et al. 2009) resulting from the diffuse $\gamma$-ray  emission of the galaxy. 
Similarly, NGC 253 has been reported by H.E.S.S. to be detected in a long campaign, along 2005-2008, with 119 h of good livetime, generating also a point-like 
5.2$\sigma$ source, being it the faintest detected so far, at the $0.3\%$ Crab level. In this case, an spectrum was not shown, but only an integrated flux at VHE, which appears lower than what models have predicted (e.g. Domingo-Santamar\'ia \& Torres 2005), promoting perhaps an interpretation based of a stronger convection of cosmic rays. Again, subsequent detailed papers from both collaborations will surely follow these preliminary presentations.

\section{Galactic sources}

\subsection{Transients, binaries, and microquasars}

In the footsteps of Agile astronomical telegrams, Hays et al. (Fermi) reported on two Fermi transients. 
At low galactic latitudes, multifrequency studies in radio/X-rays reveals one to likely be a FSRQ, with the other being yet unidentified. More such transients will appear as the survey continues.  

MAGIC presented the first strictly simultaneous X-ray and TeV observations of LSI +61 303, done with
with  XMM-Newton and Swift in September 2007 (Jogler et al.). 
They report a simultaneous outburst at X-ray and 
TeV energies, with the peak at phase 0.62 and a 
similar shape at both wavelengths. A linear fit to the 
strictly simultaneous X-ray/TeV flux pairs provides 
$r = 0.81+0.06 -0.21$,
showing a 
correlation which favors leptonic emission of the same electron population (Jogler et al.). 
VERITAS LC results were also presented, and found to be consistent with those of MAGIC 
(Holder et al.). 
Fermi also reported its detection of LSI +61 303 above 100 MeV and find it is also periodic at GeV energies, correlated with the orbit.
GeV and TeV emission are anti-correlated (what is understood as competition between production and absorption processes, plus geometry effects). Surprisingly, 
the GeV spectrum is well fitted by a power-law with a cutoff at several GeV (Dubois et al., see also Abdo et al. 2009). The latter means that either a second component of the emission is operative at low energy (pulsar magnetospheric emission?) or a yet  unidentified process smoothly join the TeV and GeV spectra. The existence of this cutoff is an unexpected feature in any published model of the binary. 
As for LS I +61 303, Fermi detects LS 5039 above 100 MeV and finds that it is also periodic at GeV energies, correlated with the orbit as well. 
GeV and TeV emission are also anti-correlated, with the GeV maximum found at periastron. In this case, the spectrum has not yet been shown, leaving the question of whether it would look similar or different to that of LS I +61 303. Are we really seeing the emergence of a population of pulsar-like GeV $\gamma$-ray  binaries, with GeV emission being directly a product of the pulsar magnetosphere?

Kerschhaggl et al. (H.E.S.S.)
reported on recent observations of PSR B1259-63, in 2005/6, at large orbital separations, and in its 2007 periastron passage. The 
resulting ligthcurve favors models not based only on the interaction between wind and disc (since significant emission out of the circumstellar wind is found, disfavoring proton-proton interaction) and pointing towards leptonic 
IC with stellar wind photons. 

Finally, Maier et al. (VERITAS)
 reported on observations of the earlier discovered HESS J0632+057 (see Hinton et al. 2009). The latter is one of the two H.E.S.S. unidentified source that is point-like in nature, 
Possibly associated with MWC 148, 
VERITAS found HESS J0632+057 to be variable: upper limits only were reported at ICRC from their 2006, 2008 and 2009 campaigns. This emphasizes the possibility for it to be $\gamma$-ray  binary.

\subsection{SNRs \& PWNe}

Both Fermi and H.E.S.S. (Uchiyama et al., Fiasson et al., respectively) reported preliminary results confirming the detection of 
W51, a middle-age SNR (30000 yrs) located  at 6 kpc, 
interacting with molecular clouds. Additionally, Giuliani et al. (Agile), Rodriguez et al. (Fermi), and Humensky et al. (a,b) (VERITAS)
reported on IC443, located at 1.5 kpc, also of about 30000 yrs. Combining these observations, and the earlier report by MAGIC (Albert et al. 2007b),
it can be seen that there is a 
displacement of the peak of the emission at different energies, from GeV to TeV, with the emission likely to be slightly extended
and with
no apparent emission from a nearby PWN. The higher the energy, the further away from the SNR shell the peak of the emission is found, what is consistent with diffusion of cosmic rays and their interaction with molecular clouds in the neighborhood. 

W44 is another SNR for which both Fermi (Tanaka et al. (Fermi)) 
and IACTs (Bugaev et al. (VERITAS)) presented preliminary reports, in this case, only Fermi has found a detection, that resulted to be significantly extended. 
VERITAS 
observed this region for 13 h.
to impose a flux upper limit of $\sim 2$\% of Crab.

On the younger SNRs, VERITAS (Humensky et al., 20 hours, 9$\sigma$ detection).
confirmed the HEGRA/MAGIC (e.g., Albert et al. 2007) detection of Cas A finding it as a  point-like source emitter. 

H.E.S.S. also reported the detection of SN 1006, that required a really long observation time
(130 h, Naumann-Godo et al.). Its 
emission was found to be at 1\% Crab, compatible with earlier upper limits, and 
compatible too with being generated in thin rims (X-ray detected, e.g., by XMM). Indeed, when looking at the XMM-detected rims with a TeV-sized PSF, the agreement between X-ray and TeV morphology is impressive. Still,
leptons or hadrons might be responsible.

The GeV--TeV connection has also been explored with observations of PWNe. Vela X and the Vela pulsar constitute a prime example of such investigations. The
PWN Vela X is the brightest 
flat spectrum radio component of the Vela SNR which is at a distance of about 290 pc. It is large in projection, 30 $\times$ 20 arcmin
Earlier H.E.S.S. observations found VHE from 5\% of its volume (the coocon, $< 0.80^0$). New 58 h made to search beyond, in a ring between 0.8$^0$ and 1.20$^0$ 
(Dubois et al. (H.E.S.S.)).
Fermi found the pulsar and the extended Vela X emission as well (Lemoine-Goumard et al.).
Radio, VHE spectrum and Fermi-LAT data for entire PWN suggests the presence of two distinct electron populations   --radio-emitting particles may be a relic population; higher energy electrons injected by pulsar. 

Enomoto et al. (CANGAROO), Hoppe et al. (H.E.S.S.), 
Gargano et al. (Fermi) presented interesting observations on H.E.S.S. J1708-443 (and the nearby PSR B1706-44)
H.E.S.S. found no 
excess at the pulsar position (however, pulses were detected by Fermi, and a glitch forced to renovate the ephemeris along the observations).
The source is extended at TeV (0.29$^0$), has a hard spectrum, 
(slope $-2$) and an integral flux above 1 TeV of 17\% Crab. CANGAROO found compatible upper limits. 
These observations are 
not easy to interpret.
If there is a high magnetic field around the PSR (estimations from X-rays yield about 140 microG, a factor of 10 larger than that of PSR 1826-1335, electrons would cool both by synchrotron and IC in similar times (and thus the TeV emission should be point like), and IC with only CMB should produce undetectable emission at TeV. 
One can assume the B-field falls off rapidly far from the PSR and electrons accumulate and escape over the lifetime of the pulsar. But the observed assymetry? In general, the morphology of PWNe is explained by proper motion, but here it is too slow (100 km s$^{-1}$), or by density gradient (but here the asymmetry goes in direction of larger density).

Finally, Wakely et al. (VERITAS)
 presented observations on two young (2000-10000 yrs PWNe)
G54.1+0.3 and Boomerang/PSR J2229+6114, yielding similar spectra (slopes of $\sim -2.3$) and fluxes (few percent of Crab).
The Boomerang PWN was first detected by Milagro at 6.6$\sigma$ and was then confirmed by VERITAS.  Geminga (or its immediate surrounding) is also detected at TeV energies by Milagro at 6.3$\sigma$.  In general, MILAGRO sees many spatial corellations with Fermi sources and detected sources are coincident with Fermi Pulsars (see Abdo et al. 2009, MILAGRO) and we comment on this below.

V\"olk et al. and Zirakashvili \& Aharonian
were among those analyzing whether current GeV-TeV observations of SNRs were enough to be unambiguously 
associated to a hadronic origin. The former authors studied young SNRs, presenting 
an analysis on whether the $\gamma$-ray  emission 
from the remnants of the type Ia supernovae SN 
1006, Tycho and Kepler can be the 
result of electron acceleration alone. 
The observed 
synchrotron spectra of the three remnants were used 
to determine the average momentum distribution of 
nonthermal electrons as a function of the assumed 
magnetic field strength. Then the inverse Compton 
emission spectrum against the CMB was calculated and  compared with 
the existing upper limits. The finding is that 
the expected interstellar magnetic fields (few tens of $\mu$G) substantially 
overpredict even these VHE upper limits and seemingly eliminate at least some
 phenomenological claims in favor 
of an inverse Compton $\gamma$-ray  scenario for these 
sources. 

On the other hand, Zirakashvili \& Aharonian
presented a 
new numerical model of the nonlinear 
diffusive shock acceleration and modeled
the SNR RX 
J1713.7-3946. They found that
neither the leptonic nor the hadronic origin of gamma-emission of complex SNR RX J1713.7-3946 can be excluded. 
However, the spectral shape of X-ray and $\gamma$-ray  spectra are better reproduced in the hadronic model.  
In the hadronic model, the SNR shock must be significantly modified throughout all surface in order to suppress thermal X-rays and to produce enough pion-decay $\gamma$-rays.  In addition the line X-ray emission of heavy ions must be also suppressed due to a low metallicity or an unusual ionization state of  the plasma downstream of the forward shock. Berezhko \& V\"oelk studied the same SNR, arguing that 
the regions of 
magnetic field ampliÞcation also are correlated with 
enhanced densities of accelerated nuclear particles 
and the associated streaming instabilities. In that case, a correlation of nonthermal X-ray and gamma- 
ray emission is not only possible but even to be 
expected for a hadronic emission scenario.

\subsection{Pulsars}

Lopez et al. (MAGIC, see also Aliu et al. 2008)
and Grondin et al. (Fermi)
 presented observations on the Crab pulsar and its nebula, joining an SED from GeV to TeV energies both for the pulsed and the steady emission.
In the case of the pulsar, a power-law with an exponential cutoff is the best fit to the data. Fermi founds the cutoff to be located at 
8.8$\pm$1.1+2.9-1.1, which is only barely compatible with the MAGIC determination:
17.7$\pm$2.8$\pm$5. Fermi could however reject a simpler
power law and a more complex
hyper-exponential cutoff  at a significance greater than 5$\sigma$. This
cutoff energy limits the height of the emission (to avoid absorption) to be beyond 4 / 6 stellar radius, what means that the polar
cap model is ruled out for the Crab pulsar.

Celik et al. (Fermi) presented results on Geminga. Again, a power-law with an exponential cutoff at $\sim$2.5 GeV was found.
A more notable change in the P1/P2 ratio with increasing energy was observed. 
For Geminga, only
upper limits were imposed  at higher energies (e.g., Finnegan et al. (VERITAS), Acharya et al. (HAGAR)).

Fermi presented several results on the discovery of GeV pulsars both in blind searches (Saz-Parkinson et al., see also Abdo et al. 2008, 2009a,b)
and after folding on the radio ephemeris (Razzano et al.). As an example, one can focus on the GeV pulsars in the Cygnus region:
each of the pulsar coincides with formerly unidentified
EGRET sources, 
only one pulsar was known from radio: 2021+365, whereas the
others are discovered by Fermi in a blind search. As these, 
more than a dozen pulsars are reported to be discovered blindly.
Other young pulsars (known from radio) were also found to be $\gamma$-ray 
bright unidentified EGRET sources, which in turn, were almost always found to be $\gamma$-ray  
pulsars in Fermi. 

\subsection{Surveys and discoveries associated to them}

Chaves et al. (H.E.S.S.) presented the general results of the H.E.S.S. Galactic plane survey extension. The
earlier H.E.S.S. survey covered the longitude range $\pm$30$^0$ from the Galactic center, and the latitude range $\pm$3$^0$, and implied 95 h plus follow up  observations up to a total of 230 h, with 28 min duration pointings with 0.70$^0$ offsets. The
current H.E.S.S. survey covers longitude in the range 275$^0$ -- 60$^0$, and latitudes less than 5$^0$, and involves  1400 h of data, 450 h of which were taken in survey mode. Several of the sources mentioned, and others that we discussed below (e.g., W51 / 1741-302 / 1708-443 / 1505-622 / Westerlund 1 region) were discovered in this survey.
Tibolla et al. (H.E.S.S.)
 presented some of the sources in more detail:

H.E.S.S. J1507-622: 
 It is a bright source, at 8\% of Crab, located 
 at 3.5$^0$ off the plane. It is
 extended, 0.11$^0$ and has a hard spectral slope ($-2.2$), with 
 no obvious counterpart

H.E.S.S. J1503-582 
 It is a bright source, at 6\% of Crab, also with no
 obvious counterpart (a PSR with low $\dot E$,  a cataclysmic binary nearby
 a Forbidden Vel. Wing?).
 It is also extended.

H.E.S.S. J1848-018:
 It is a source at 2\% of Crab, with no
 obvious counterpart (although nearby the star forming region W43, and the star WR121a).

Near the Galactic center several sources have been found:
 
 H.E.S.S. J1745-303 
 With plenty of possible counterparts: 2 SNRS, 1 EGRET source, 1 PWN, molecular clouds

H.E.S.S. J1741-302
 At 1\% of crab, at the lower end of H.E.S.S. sensitivity, with 
 two apparent hot spots (although the claim is that current statistics is not enough for source morphology).
 Pulsars in the region have low spin down, so that counterparts are not identified.

This is in addition to the Galactic center source, which is
H.E.S.S. J1745-290 (see  Vivier et al. (H.E.S.S.)) and for which the 
 latest dataset comprises 2004-2005-2006 campaigns, and 93 h of observations. For the first time, 
 a deviation from pure power law is visible (last point reaches up to 70 TeV). 
 No flares, no correlation with X-rays were found.

As one can see, all sources appear extended with no obvious counterpart:
Why are there so few point-like VHE sources? Is this the telltale of an evolutionary effect on $\gamma$-ray  binaries, being them the only point like emitters?

Ohm et al. (H.E.S.S.)
and de O\~na Wilhelmi et al. (H.E.S.S.)
reported VHE observations on the interesting clusters Westerlund 1 and 2. 
Westerlund 1 is currently the record holder in terms of its population of stars in the Wolf-Rayet (WR) phase. At least 24 WR stars are known of which 60-70\% are expected to be in binary systems. Moreover, more than 80 blue super-giants, 3 red super-giants, and one luminous blue variable are known to exist there. The Westerlund 1 region was seen by H.E.S.S.  in VHE $\gamma$-rays, after observing 34 h livetime, with a 
680 GeV threshold. Possible acceleration sites in the region are plenty: colliding wind binaries, collective winds, supernovae, but none of them are individually identified. In the case of Westerlund 2, after following an indication of a possible energy dependent shape that led to re-observations, 
a new source was discovered, coincident with a Fermi-LAT
detection associated with a pulsar (top 20 in $\dot E/d^2$): The region is now known to host two nearby GeV -- TeV sources.

Additional surveys were presented at the ICRC 2009. Most notably, the VERITAS effort reported on its survey of the Cygnus region (Weinstein et al.). The Cygnus survey covers, 
between 67 and 82 degrees in galactic longitude and between -1 and 4 degrees in galactic latitude
and has more than 140 hours of observations (112 h in based survey, $\sim$30 more hours follow up), with an average sensitivity of 
4\% Crab. No hotspots above 5$\sigma$ post-trials in the base survey were found, implying a
99\% CL point-source flux limits (for all points in
survey below 3$\sigma$) are $\sim$3\% of the Crab above 200 GeV or
8.5\% Crab UL,  if extended. Note that the nearby HEGRA/MAGIC unidentified source TeV J2032+4130
(e.g., Aharonian et al.  2002, Albert et al. 2008) has a detected flux that is less than the obtained coverage in the VERITAS survey. 

At even higher energies, MILAGRO also presented the results of its sky survey at the ICRC (Smith et al., Pretz et al.). Out of the 
205 sources in the Fermi bright source list, 
34 of them are not identified with extragalactic objects, and 
14 of them are correlated with MILAGRO excesses (this is at more than 5$\sigma$ off of what is expected out of a random correlation). All of the brightest MILAGRO sources are spatially coincident with Fermi pulsars, a correlation which final physical reason is yet unclear.

\section{Extragalactic Sources}

The ICRC 2009 enjoyed many reports on the discovery of high and VHE emission from radiogalaxies.
Among the most important, I can quote the several reports on the radio-TeV multiwavelength campaign on M87, e.g.,  by
Hui et al. (VERITAS), Tescaro et al. (MAGIC), and Wagner et al (H.E.S.S., VERITAS, MAGIC, VLBA).
M87, the central galaxy of the Virgo
Cluster, was the first radio galaxy detected in the TeV
regime (Aharonian et al. 2003, 2006). With a jet structure off from the line of sight, and that is spatially resolved in X-ray
(by Chandra), in optical and in radio observations, until the present report, 
the TeV emission has not yet been well localized. In 2008, the three main atmospheric
Cherenkov telescope observatories (H.E.S.S.,
MAGIC, and VERITAS) coordinated their observations
in a joint campaign from January to May with a
total observation time of 120 hours. The campaign
largely overlapped with an intensive VLBA project
monitoring the core of M87 at 43 GHz every 5 days.
In February, high TeV activities with rapid flares
have been detected. Contemporaneously, M87 was
observed with high spatial resolution instruments in
the X-ray band (Chandra). 
Simultaneous Chandra observations found the central knot HST-1, the
innermost knot in the jet, in a low state, while the
nucleus of M87  showed increased X-ray activity that was correlated with the TeV flare observed by all three telescopes participating in the campaign, and additionally, showed increased radio emission as observed with the imagining capabilities of VLBA. 
This consistent increase of radio, X-ray, and TeV emission, with the first two being well localized at the nucleus of M87, suggest the the origin of the latter one is localized therein as well. 
This is in
contrast to the 2005 VHE $\gamma$-ray  flare, where HST-1 was
in an extreme high state. In the light of the 2008 results, a model suggesting
HST-1 to be the origin of the $\gamma$-ray  emission seems
less likely  (the caveat is that similar VLBA radio coverage of HST-1 was not obtained, see Acciarri et al 2009, for further details).

An additional radiogalaxy presented at ICRC was Cen A (Lenain et al. (H.E.S.S.)), observed (115 h) at very low flux level, of only 0.8\% Crab, with a position compatible with radio core and inner (kpc) jet
and a power law slope of $-2.73$ fitting the spectrum.

Reyes et al. (VERITAS, Fermi) and  Errando et al. (MAGIC)
presented interesting reports on a third radiogalaxy candidate 3C66B, which happens to be close to a blazar. In fact, 3C 66A and 3C 66B are two AGNs separated by 6 arcmin in the sky. 
3C 66A is a distant blazar with redshift of 0.444 (uncertain). 
3C 66B is a large Fanaroff-Riley-I-type radio galaxy, similar to M87, with redshift of 0.0215.
The 
MAGIC source reported (observed in 2007, Aug-Dec) is 
	incompatible in position with 3C66A,
	active at 2 TeV, and presenting 
	no significant variability
The 
VERITAS/Fermi source reported is instead 
	located at the 3C66A position, and 
	observed in a strong flare in 2008, Oct., thus a year after. Although the Fermi localization clearly favors blazar 3C 66A as the $\gamma$-ray  source counterpart, some small contribution from radio galaxy 3C 66B cannot be excluded either.
The different localization and spectra ($-3.1$ for the MAGIC source, $-4.1$ for the VERITAS one) as well as the year that has passed between the two observations point to the possibility that both sources could indeed be real TeV emitters. 

Several new blazars have been presented at the ICRC. I will summarize these discoveries in what follows.
The first notable such case is W Comae (Maier et al. (b) (VERITAS)). W Com is an intermediate-frequency-peaked BL Lac (IBL) at a redshift of z = 0.102.
W Com was discovered during a strong outburst in March 2008 with a second flare  on June 7 and 8, 2008 with a three times higher flux at $E>$200 GeV. Its SED presents a too low magnetic field for single SSC models and it is thus
ideal for MW campaigns, with simultaneous study of flares, that will likely allow for model distinction

Perkins et al. and also Benbow et al. (VERITAS)
also presented the discovery of two typical HBL AGNs, 
1ES 0806+524 \& RGB 0710+591 and one
interesting IBL:  PKS 1424+240. The latter was the first VHE discovery motivated by Fermi.
In this case, the VHE finding followed up a detection by Fermi LAT, and  was taken up
by both VERITAS (ATEL 2084) and MAGIC (ATEL 2098).
H.E.S.S. presented another new HBL: RGB 0152+017 at $z=0.08$
(Kaufmann et al. (H.E.S.S.), Zech et al. (H.E.S.S.)), with a spectrum fitted by a power law with slope $-3.15$.
MAGIC followed up an optical trigger to find S5 0716+714, which is just the third low frequency BL Lac (Mazin et al. (MAGIC)).

The latter discovery prompts to put in context the optically triggered VHE observations (Lindfors et al. (MAGIC), Barres de Almeida et al.) . 
Usual optical triggers use only photometry, and this 
could be problematic when optical emission comes from larger regions (similar to X-rays).
Still, it is unclear whether the link of optical triggers and VHE is always (or when, if not always) fulfilled. 
Why is the optical-VHE link still unclear? To prove the connection one needs a correlation study.
But good enough VHE $\gamma$-ray  light curves exists only for 2-3 sources and 
VHE $\gamma$-ray  data is also too sparse and covers too short periods of time to get statistics on duty cycles of the sources. One needs to assume the duty cycle: Should one assume (e.g.) Mrk421 is a good representative of the whole population? Or should one assume that VHE $\gamma$-ray  duty cycle is similar to optical duty cycle of the source (albeit in some cases, e.g., Albert et al. 2007, no significant correlation between gamma-rays and optical data is seen)? This inherits lots of biases that make the study doubtful. There is an interesting seeming correlation between flares
in HE and VHE energies and an increase in polarization, although this correlation is not yet used as a triggering technique. 
At lower energies (Behera et al.) found that using a 10 blazars sample an apparent dichotomy between 
optically-faint / GeV bright ($z>0.36$) and optically / 
GeV faint ($z<0.158$) exists. Is this
maintained in a large sample?

Of course, the advent of Agile and Fermi has shown up in exquisite AGN observations at the ICRC. An example of this are the 7 months of observations of 3C279 by Fermi, with a clear detection of low and high states, and daily lightcurves (Hayashida et al.) or a 2 days bins determination of flux and spectrum along 6 months of observations of Mrk 421 
(Paneque et al.). Agile similarly reports on long campaigns, like e.g., on 3C 454.3 (Marisaldi et al.).
Agile is finding previously-known AGNs yet, consistent with the most luminous
Fermi detections. From Fermi results one can see that all types of blazars has been detected: quasi simultaneous SEDs for 48 LBAS
were shown by Mazziotta et al. (a, b) (Fermi):
Not all consistent with power laws (e.g., see 3C 454.3) and not all well fitted with SSC models (especially for LBLs).

New campaigns on old friends, like Mrk 421 have also been presented. For Mrk 421, a MAGIC - VERITAS - Agile - GASP-WEBT campaign on the June 2008 flare has been shown, with
SSC models in two simultaneous periods of observations favoring variability as a result of changes in the electron distribution, see
Wagner et al. and Bonnoli et al. (MAGIC) for details.
Grube et al. (VERITAS)
presented flux and spectral variability on short timescales
(but with coverage up to 1 month before Fermi launch).
PACT, HAGAR, ARGO-YBJ
campaigns were also available
(e.g., Chitnis et al.).

For 3C279, observations reported included those made in 2007: 8 nights of observation with MAGIC, that detected the object in only one night of those, with a hard spectrum compatible with 2006 first detection (Berger et al., Aliu et al. 2008b). This re-detection is an important confirmation for such a distant AGN.

Another interesting campaign was that of PKS 2155-304, with observations during 11 days, finding flux and spectral variability (Gerard et al. (H.E.S.S., Fermi)).
A 1 zone SSC model can reproduce the SED
but is not at ease with the correlation pattern: If the same population of electrons drives these emissions then the optical/HE and HE/VHE ßuxes should also be correlated, which does not appears to be the case.

Finally, the Whipple 10-m telescope continues to do monitoring (on every moonless night the sources are visible) on Mrk 421, Mrk 501, H 1426+428, 1ES 1959+650, and 1ES 2344+514, so that if detection of a flare proceeds, alerts to VERITAS can be submitted. This monitoring program is indeed creating long-term lightcurves with spacing that are usually unavailable from other instruments (Pichel et al.). See Hsu et al. for the MAGIC efforts in this sense.

\section{GRBs}

The
brightest burst that LAT observed up to now, GRB 080916C, was reported at the ICRC by Tajima et al. (see also Abdo et al. 2009).
At $z=$4.35,  the redshift implies the largest (apparent) isotropic
$\gamma$-ray  energy release. 
A single spectral form that fits the energy range from 8 KeV to 
13.2 GeV is possible; observations available at such low energies because of the simultaneous use of the GBM, also onboard Fermi.
The LAT can be used as a counter to maximize the rate and to study time structures above tens of MeV.
More than 300 LAT photons in the first 100 seconds were detected showing that
the first low-energy peak is not observed at LAT energies.
Spectroscopy needs LAT event selection ($>100$ MeV) and
5 intervals for time-resolved spectral analysis were chosen 0--3.6--7.7--16--55--100 s.
14 events above 1 GeV were recored. The analysis shows that the 
$E>100$ MeV emission is approximately 5s later than the
$E<1$ MeV emission. This trend appears to be common to other bursts.

Omodei et al. (Fermi) presented further examples of GRBs detected by the LAT, including the first GRB measured but it: 080825C. For this burst too, emission in the 100 MeV regime is
apparently delayed (or correlated with a second peak at low energies), although  few events limits statistical confidence in many 
observed bursts.
 In total
there have been, at the time of the ICRC, 241 GBM  detections, and only
9 LAT detections 
(out of 129 in FoV).
Razzaque et al. proposed that the GRBs accelerate protons and they emit synchrotron radiation, thus the time needed to accelerate particles in order for them to emit at the LAT range could explain the delay.

Searches for GRB emission at higher energies have been thoroughly done and presented as well, all yielding to upper limits.
MAGIC and also VERITAS attained a few minutes of delay for several bursts, see
Galante et al. (VERITAS), Garczarcyk et al. (MAGIC), Covino et al. (MAGIC), Gaug et al. (MAGIC). In some cases, e.g, GRB 090102,
MAGIC attained an upper limit at only 50 GeV. At even higher energies, perhaps the most interesting upper limit is that on
GRB 080319B (Aune et al. (MILAGRO)).
GRB 080319B was dubbed the naked-eye gamma- 
ray burst, as an observer under dark skies could 
have seen the burst without the aid of an instrument. 
This burst benefit from the space and time proximity to GRB 080319A, that allowed 
prompt emission from GRB 080319B to be detected by several wide-Þeld instruments.
Milagro observed 
GRB 080319B and no signiÞcant gamma-ray signal was 
detected with either the scaler (1 -- 100 GeV, where the single hit rates of all 
of the Milagro PMTs were recorded once a second) or standard 
(50 GeV -- 100 TeV) analyses.

Stacking analysis of several GRBs also yielded to upper limits (see, e.g., Guo et al. (ARGO-YBJ), DiGirolano et al. (ARGO-YBJ), and the interesting upper limits from ICECUBE in the neutrino domain by Meagher et al.,  Duvoort et al.)

\section{Astrophysical objects we have not yet detected at high or very-high energies}

Many, and on a varied set of different astrophysical objects, have been the upper limits reported at the ICRC 2009. For most of these cases, permanent fluxes of a few percent of Crab are ruled out. A summary of such objects (or classes of objects) is made in Table 1, together with the corresponding references.

  \begin{table*}[t]
  \caption{Astrophysical objects not yet detected at high
  and very-high energies}
  \label{table_simple}
  \centering
  \begin{tabular}{ll}
  \hline
   Object or class  & Reference \\
   \hline 
    
 Clusters & \ldots \\
Abell 3376 & Galante et al. (VERITAS) \\
Abell 3376 & Matoba et al. (CANGAROO) \\
NGC 1275 & Galante et al. (VERITAS) \\
Coma & Galante et al. (VERITAS) \\
Abell 85 & Domainko et al. (H.E.S.S.) \\
    \hline
    Radiogalaxies and elliptical galaxies & \ldots\\
    3C 11  & Galante et al. (VERITAS) \\
    In the Virgo, Coma and Fornax fields & Pedaletti et al. (H.E.S.S.) \\
    \hline 
    Extended emission around blazars   & Sitarek et al. (MAGIC) \\
    \hline
    Binaries & \ldots \\
Several   LM and HMXRBs & Guenette et al. (VERITAS)\\
 Several  LM and HMXRBs &   Saito et al. (MAGIC) \\
    GRS 1915+105 & Szostek et al. (H.E.S.S.) \\
    Wolf-Rayet binaries & Torres et al. (MAGIC) \\
    \hline
    Globular clusters & \ldots \\
    M13 & Jogler et al. (b) (MAGIC) \\
    M13, M15, M5 & McCutcheon et al. (VERITAS) \\
    \hline
    Magnetars & \ldots \\
    at high energy & Vasileiou et al. (Fermi) \\
    at very-high energies & Guenette et al. (b) (VERITAS) \\
    \hline
    Pulsed radiation from PSRs (not Crab) & de los Reyes et al. (MAGIC) \\
\hline
SNRs & \ldots \\
Tycho &     Carmona et al. (MAGIC) \\
Radio selected SNRs & Carmona et al. (MAGIC) \\
Forbidden velocity wing structures & Holder et al. (b) (VERITAS) \\
\hline
Any individual source above 100 TeV & Feng et al. (TIBET)\\
 \hline
  \end{tabular}
  \end{table*}

\section{Dark matter}

There was a single contribution focusing on dark matter studies in the ICRC sessions reported in this summary (Ripken et al. (H.E.S.S.)), comparing the latest results on the observations of the Galactic Center at TeV energies with predictions of models
including radiative corrections. The conclusion was that dark matter can have a minimal, if at all, contribution 
to the radiation observed for most of the phase space explored. 

\section{Unconfirmed claims}

Many claims were made by the SHALON experiment in posters presented at the ICRC 2009 (continuing a series of presentations in previous meetings). Among them: morphology, spectra, and variability detections were claimed in a series of papers by  Sinitsyna et al. on 
Cyg X3, 
Tycho, 
Geminga, 
NGC 1275 and SN 2006gy, in the central part of Perseus,
3C54.3 (z=0.859),
1739+522 (z=1.375), and
Crab. Neither of these claims were confirmed by observations made (sometimes concurrently) with other --far more sensitive and larger-- experiments, although that fact is not carefully assessed in any of the papers quoted.

\section{Concluding remarks}

\subsection{Diffuse emission} 

\begin{itemize}
\item The GeV bump that led to hundreds of papers was 
not confirmed, no GeV excess is visible in Fermi data.
\item No dark matter, no 
fundamental change in understanding of CRs in the 
galaxy appears to be necessary. 
This emphasizes the fact that differentiating $\gamma$-ray  
sources from the diffuse component is the key of the field. 
All GeV knowledge lives and dies with how well we 
understand the diffuse 
\item Extragalactic star forming regions are $\gamma$-ray  
sources. This discovery is very important and necessary to check our 
understanding of CR production and propagation. 
Be careful: The LMC at VHE and HE looks different at
current sensitivity. 
\item The conversion between CO and H2, known as the X constant is, well... variable (see 1st point), depending on location.
\end{itemize}

\subsection{Transients, binaries, and microquasars} 

\begin{itemize}
\item LS I +61 303  never stops to surprise us...
The discovery of a cutoff in the GeV spectrum of LS I +61 303 is one of the most unexpected results of this conference
What does it really mean? How does it relate with orbital variability? No model has predicted this cutoff.
X-ray / TeV correlation looks solid, better statistics would not hurt.
\item Gamma-ray binaries are yet to be understood... more candidates are necessary, HESS J0632+057 is promising
\end{itemize}

\subsection{SNRs \& PWNe}

\begin{itemize}
\item Middle-age SNRs (IC 443, W44, W51É) provides excellent laboratories to study cosmic-ray propagation and diffusion, they can shine both at TeV and GeV, with co-located or dis-located positions, steep or hard spectra, according to diffusion parameters, distance to target, etc.
We are just starting to measure CR propagation, but still a difficult task: we need to identify the GeV and TeV sources to be related to the SNR (directly or indirectly).
Extension helps, especially if there is no competing interpretation (a PWN), or if such can safely be ruled out.
\item Unambiguous evidence for the origin of cosmic-ray at ICRC 2009? Although hadronic models are favored for young SNRs, discussion is still ongoing.
\end{itemize}

\subsection{Pulsars}

\begin{itemize}
\item Amazing spectra has been shown joining GeV and TeV energies: this has happened here for the first time in history (e.g., see the cases of Crab pulsar and Crab nebula)
\item Data now at hand implies the outer gap model is essentially correct (not a single pulsar spectrum shown is better fitted by polar caps)
\item Fermi results on pulsars are to provide both single and population knowledge: fulfilling and exceeding expectations.
Most brightest Galactic EGRET unids are Fermi pulsars.
Fermi science (at its 1-year sensitivity level) is just starting.
\end{itemize}

\subsection{Surveys and discoveries associated to them}

\begin{itemize}
\item General indication of different $\gamma$-ray  source density: less sources in the northern sky. The
density of H.E.S.S. survey: 12 sources in the inner l=30$^0$ with more than 5\% Crab.
In the VERITAS survey, at equal density, this would imply $\sim$3 sources, when 0 have been detected. 
\item A MILAGRO -- Fermi connection was reported but yet to be understood (it is key for our global understanding of gamma-emission).
All of the brightest MILAGRO srcs are spatially coincident with Fermi pulsars. Are the latter especial in any way within the Fermi population? Does not seem so at first sight. Why are there many other similar Fermi sources with no counterparts in the MILAGRO map?
\item Westerlund 1 and 2 are key sources: need to distinguish how (and whether) the star forming regions as a whole are related to them.  
\end{itemize}

\subsection{Extragalactic Sources}

\begin{itemize}
\item MW still drive the field to wonderful results,
although 
have to be strictly simultaneous to make real sense. 
From these campaigns: Nothing strikes me as being completely and irremediably at odds with the current understanding of the different kind of blazars and the influence of internal and external fields in the $\gamma$-ray  production. 
The Fermi -- TeV observatories joint studies are just starting up, and is to provide a comparison framework for hadronic-leptonic models.  
\item HBLs, IBLs, LBLs all are $\gamma$-ray  sources
\end{itemize}

\subsection{GRBs}

\begin{itemize}
\item Fermi (LAT and GBM) impact on the GRB field is bound to be large.
Already obtained results point to a general delay of $\gamma$-ray  emission with energy.
\item VHE experiments are still awaiting to catch a GRBs, improvements in reaction time, sensitivity, and threshold renders a reasonably optimistic  expectation (if there is emission much beyond the Fermi range).

\end{itemize}

\section*{Acknowledgments}

I would like to express my thanks to the organizers for inviting me to give the Rapporteur talk and write this summary of it. I also thank 
the collaborations (especially H.E.S.S. and VERITAS) and to R. Ong and W. Hoffman in particular,
who have given access to the results presented here a bit 
beforehand to facilitate the presentation at the meeting. Finally, I thank the many colleagues who have send in comments and patiently involved me in discussions at the ICRC. I thank Olaf Reimer for a critical reading of the manuscript. 
This work has been supported by grants AYA2006-00530, AYA2008-01181-E/ESP,
and SGR2009-811.

\end{document}